%% file: emnlp2021.tex
\pdfoutput=1

\documentclass[11pt]{article}

\usepackage[]{emnlp2021}

\usepackage{times}
\usepackage{latexsym}

\usepackage[T1]{fontenc}

\usepackage[utf8]{inputenc}

\usepackage{microtype}

\usepackage{booktabs} 
\usepackage{upgreek} 
\usepackage{tikz-cd}
\usepackage{tikz}
\usetikzlibrary{shapes,arrows}
\usepackage{hyperref}
\usepackage{graphicx,verbatimbox}
\usepackage{listings}
\usepackage{amsmath}
\usepackage{subcaption}
\usepackage{mathtools}
\usepackage{color}
\usepackage{tabularx,ragged2e}
\usepackage{multirow}
\usepackage{mathtools}
\usepackage{enumitem}
\usepackage[british]{babel}
\graphicspath{{images/}}

\title{Dealing with Typos for BERT-based Passage Retrieval and Ranking}

\author{Shengyao Zhuang \\
  The University of Queensland \\
  Brisbane, QLD, Australia \\
  \texttt{s.zhuang@uq.edu.au} \\\And
  Guido Zuccon \\
  The University of Queensland \\
   Brisbane, QLD, Australia\\
  \texttt{g.zuccon@uq.edu.au} \\}

\begin{document}
\maketitle
\begin{abstract}
Passage retrieval and ranking is a key task in open-domain question answering and information retrieval. Current effective approaches mostly rely on pre-trained deep language model-based retrievers and rankers. These methods have been shown to effectively model the semantic matching between queries and passages, also in presence of keyword mismatch, i.e. passages that are relevant to a query but do not contain important query keywords. 

In this paper we consider the Dense Retriever (DR), a passage retrieval method, and the BERT re-ranker, a popular passage re-ranking method. In this context, we formally investigate how these models respond and adapt to a specific type of keyword mismatch -- that caused by \textit{keyword typos} occurring in queries. Through empirical investigation, we find that typos can lead to a significant drop in retrieval and ranking effectiveness. We then propose a simple typos-aware training framework for DR and BERT re-ranker to address this issue. Our experimental results on the MS MARCO passage ranking dataset show that, with our proposed typos-aware training, DR and BERT re-ranker can become robust to typos in queries, resulting in significantly improved effectiveness compared to models trained without appropriately accounting for typos.

%
\end{abstract}

\input{sections/introduction}
\input{sections/method}

\input{sections/experiment}

\input{sections/results}

\input{sections/case_study}
\input{sections/conclusion}

\section*{Acknowledgements}
Dr Guido Zuccon is the recipient of an Australian Research Council DECRA Research Fellowship (DE180101579). This research is partially funded by the Grain Research and Development Corporation project AgAsk (UOQ2003-009RTX).

\bibliography{anthology,custom}
\bibliographystyle{acl_natbib}

\appendix

%

\end{document}

%% file: sections/introduction.tex
\section{Introduction} \label{intro}
Passage ranking is a core task for many information retrieval related applications. In the context of conversational search and question answering, for example, passage ranking is often the first step in the system's pipeline: thus the quality of the ranking results will affect the effectivenesses of the downstream tasks. 
Traditional passage ranking models, such TF-IDF and BM25, use exact keyword matching signals, where a retrieved passage must contain at least one of the query's keywords. This mechanism however limits the capability of these models to retrieve passages that are semantically relevant but use different keywords: this is the well-known vocabulary mismatch problem.

Recent advances in NLP have seen the introduction of deep language models~\cite{devlin2018bert,brown2020language,raffel2020exploring}; BERT~\cite{devlin2018bert} in particular has shown generalised promise in language understanding tasks. BERT adopts the transformer encoder~\cite{vaswani2017attention} as model architecture and uses WordPiece token embeddings~\cite{wu2016google} as model inputs. This design allows BERT to deal with the vocabulary mismatch problem.
Hence practitioners have turned to design BERT-based passage ranking models~\cite{lin2020pretrained}.

Two main directions have been adopted to exploit BERT for effective passage ranking:

\begin{itemize}[noitemsep,nolistsep,leftmargin=*]
		\item \textbf{Dense Retriever (DR)}~\cite{zhan2020repbert,xiong2020approximate,gao2020complementing,khattab2020colbert,karpukhin2020dense,ding2020rocketqa,luan2020sparse}: queries and passages are separately encoded into low-dimensional dense representations with BERT. At indexing time, passage representations are computed and then stored in the index. At query time, a single query encoder inference is needed to obtain the query representation; then passage relevance scores are estimated by computing the similarity between the query and passages' representations.
		\item \textbf{BERT re-ranker}, a.k.a. monoBERT
		~\cite{nogueira2019passage,dai2019deeper,gao2021rethink}: the ranking task is modelled as a classification task that builds upon the BERT model. The input to BERT is a $<query,passage>$ pair and the relevance score can be computed by a linear layer on the $<CLS>$ token embedding, or the query likelihood estimated by the BERT model~\cite{zhuang2021tilde}. A key drawback of BERT re-ranker models is that multiple inferences are required at query time: this is a computationally expensive process which results in high query latency~\cite{zhuang2021fast,macavaney2020expansion,hofstatter2020interpretable}. Thus the use of these methods is confined to second stage re-ranking. 

\end{itemize}

In principle, these methods are not affected by keyword mismatch because they use the latent embedding space to estimate the relevance of a query to a passage. This is supported by recent work that has shown the DR and BERT re-ranker provide better semantic matching~\cite{zhan2020repbert,macavaney2020expansion,formal2021white}.

In this paper we investigate the impact of a specific type of keyword mismatch: that caused by the presence of typos in the query. 
Traditional exact keyword matching methods perform badly on queries that contain typos. Extra query processing steps, such as spelling correction, are required for these methods to be tolerant to typos in queries~\cite{martins2004spelling}. On the other hand, it is expected that BERT-based models can handle typos occurring in queries well. This is because BERT uses the WordPiece algorithm which splits a keyword that does not match an entry in the BERT vocabulary (typos are likely to not be present in this vocabulary) into character-level sub-tokens. This can be used to produce embeddings for out-of-vocabulary keywords, which are then passed as input to the BERT encoder. However, this intuition has never been tested before, and the capability of BERT-based passage ranking models to deal with typos in queries has not been quantified.

To address this gap, we first formally investigate how the BERT-based DR and re-ranker respond and adapt to queries that contain typos. Specifically, we use different typo generators to produce typos for queries; we then compare the effectiveness of the rankers when using queries with typos vs. without typos.
Interestingly, we find that these models fail to handle queries with typos -- typos can lead to a significant drop in effectiveness for both DR and BERT re-ranker. In order to solve this issue and obtain typo-robust ranking models, we then propose a simple typos-aware training strategy, in which queries with typos are produced and used also for training. Our experimental results on the MS MARCO passage ranking dataset show that, with our typos-aware training, DR and BERT re-ranker can become robust to typos in queries, without loss in effectiveness for queries without typos.


%% file: sections/method.tex
\section{Methodology} \label{methodology}

With respect to BERT-based models for passage retrieval and typos in queries, we investigate the following research questions:

\begin{itemize}[noitemsep,nolistsep,leftmargin=*]
	\item \textbf{RQ1:} What is the impact of typos in queries on BERT-based DR and re-ranker effectiveness?
	\item \textbf{RQ2:} Do different typo types affect the effectiveness of the BERT-based methods differently?
	\item \textbf{RQ3:} Does the proposed typos-aware training improve the effectiveness of the BERT-based methods on queries with typos? Does it hurt their effectiveness on queries without typos?
\end{itemize}

\subsection{Synthetic Typo Generation } \label{sec-typos-generation}
To answer our research questions, a reasonably large set of queries with different types of typos is needed. As there is no available dataset for passage retrieval with labels that indicate the presence of typos in queries, we set off to create one such dataset. For this we augmented the MS MARCO passage retrieval dataset. (Manual inspection of this dataset did not reveal a considerable amount of queries with typos; the dataset curators likely did manually remove most typos). For augmentation, we synthetically generated typos from the original queries in the dataset, so that we could carefully control the number and types of typos. 


For generating typos, we used the following operations that give rise to typos that often occur in real-world queries~\cite{hagen2017large}:

\begin{itemize}[noitemsep,nolistsep,leftmargin=*]
	\item \textbf{Random character Insertion (RandInsert):} Inserts a random letter into a random word, e.g., ``search typo'' -> ``search tyapo''.
	\item \textbf{Random character deletion (RandDelete):} Deletes a random character of a random word, such as ``search typo'' -> ``search tpo''.
	\item \textbf{Random character substitution (RandSub):} Randomly replaces a character of a random word with a random letter, e.g., ``search typo'' -> ``search type''.
	\item \textbf{Swap neighbor character (SwapNeighbor):} Randomly swaps a character with one of its neighbor characters, e.g., ``search typo'' -> ``search tyop''.
	\item \textbf{Swap adjacent keyboard character (SwapAdjacent)}: Randomly swaps a character with one of its adjacent letter on the keyboard\footnote{ e.g., on a \textit{QWERTY} keyboard, the list of adjacent characters for character `s' is [`q', `w',`e', `a', `d', `z', `x', `c'].}
	, e.g., ``search typo'' -> ``search typi''.
\end{itemize}
Since queries in MS MARCO are relatively short ($\approx6$ keywords on average)~\cite{nguyen2016ms}, when generating typos for queries, we only consider keywords that have more than 3 characters and only randomly modify one keyword per query. We use open-source tool kits TextAttack~\cite{morris2020textattack} to implement these typo generators.

\subsection{Typos-aware Training}
To deal with queries with typos we propose to consider such queries also during the training phase of DR and BERT re-ranker: we call this typos-aware training. Specifically, for each original query that appears during the training phase, we draw an unbiased coin. If the result is head, we leave the query unchanged and use it for training. If it is tail (50$\%$ chances) we inject a typo in the query by uniformly sampling one of the considered typos generators (Section~\ref{sec-typos-generation}), and use the modified query for training. By doing so, at training time, the BERT-based methods will observe  both the original, typos-free, queries and queries with different types of typos. Thus, in order to reduce the training loss, we force the methods to learn to be invariant to different types of typos. 


Our typos-aware training can be considered a data augmentation approach, with small perturbations to some training queries: these do not change the underlying intent of the query or the relevance of the target passage.  Data augmentation has been shown effective for a range of deep learning tasks, including computer vision~\cite{he2020momentum,chen2020simple,grill2020bootstrap} and NLP~\cite{zhang2015character,wei2019eda,xie2020unsupervised,jiao2020tinybert}; however, the impact of data augmentation on ad-hoc retrieval remains to be studied.


%% file: sections/experiment.tex
\section{Experimental Settings} \label{results}
\subsection{Dataset and Evaluation Measures}
For evaluation, we use the MS MARCO passage ranking dataset~\cite{nguyen2016ms}, which consists of 8.8M passages, $\approx$503K training queries and 6,980 dev queries. For typos-aware training, we modify training queries with a 50\% chance. For dev queries, we experiment with both the original queries and with queries modified to contain typos. We produce typos for all queries in the dev set using the strategies in Section~\ref{sec-typos-generation}; we also consider the average effectiveness across typos queries.

We use the official metric MRR@10 to evaluate the ranking effectiveness of both DR and BERT re-ranker. We use the BERT re-ranker as a second stage ranker, on top of the initial rankings provided by DR. This is unlike previous work~\cite{lin2020pretrained}, in which the BERT re-ranker is typically used on top of BM25. Our setting is motivated by the fact that BM25 would fail to retrieve the relevant target passages for queries that contain typos. Because of this, we also report Recall@1000 (labelled Recall) for DR,  as this forms the basis of the first stage of retrieval and the number of retrieved relevant passages affects the effectiveness of the BERT re-ranker. Recall for BERT re-ranker is thus the same as that of DR, and is not reported.

\subsection{DR Training Details}
We follow~\citet{zhan2020repbert} when training the DR.  We adopt the BERT-Siamese architecture in which the query encoder and passage encoder share the BERT model parameters. This architecture has been used consistently in many recent approaches~\cite{luan2020sparse,xiong2020approximate}. We use pairwise hinge loss with the "Train Triples" data provided in MS MARCO  to fine-tune the ``bert-base-uncased" model from the Huggingface library~\cite{wolf-etal-2020-transformers}. We use the ADAM optimizer, learning rate of 3e-6 with linear warm-up and decay scheduling. The model is trained on a single Tesla V100 GPU with a batch size of 26 and gradient accumulation step of 2 for 210K steps.

\subsection{BERT re-ranker Training Details}
To train the BERT re-ranker, we follow the training practice described by~\citet{nogueira2019passage}. We fine-tune a ``bert-large-uncased" model with binary cross-entropy loss to perform binary classification on query-passage pairs. Negative pairs are randomly sampled from the top 1,000 passages retrieved by a trained DR model (without typos-aware training). We set the ratio of positive pairs to negative pairs to 1:4. We use the same optimizer and learning rate scheduling used for DR; the model is trained on two Tesla V100 GPUs with a batch size of $2 \times 64$ for 70K steps. 

For both DR and BERT re-ranker typos-aware training, we use exactly the same setting used for the standard training described above.

%% file: sections/results.tex
\begin{table*}[t]
	\centering
	\caption{MS MARCO passage ranking results. Row 2 reports results averaged across all typos queries; rows 3-7 results for each typos type (for each type, typos are injected in all dev queries). Percentage reductions are computed w.r.t. the original queries; bold represents best performance across training methods for each of DR and BERT re-ranker. Statistical significant gains (two-tailed paired t-test with Bonferroni correction, $p<0.01$) obtained by models with typos-aware training over the models with standard training (std.) are indicated by ${\dagger}$. }
	\resizebox{2.1\columnwidth}{!}{
		\begin{tabular}{ l | l l | l l | l l | l | l}
			\toprule
			\multicolumn{1}{c|}{}& \multicolumn{2}{c|}{\textbf{BM25}}& \multicolumn{2}{c|}{\textbf{DR (std.)}} & \multicolumn{2}{c|}{\textbf{DR (typos-aware)}} & \multicolumn{1}{c|}{\textbf{Re-ranker (std.)}}& \multicolumn{1}{c}{\textbf{Re-ranker}}\\
			
			\multicolumn{1}{c|}{}& \multicolumn{2}{c|}{}& \multicolumn{2}{c|}{} & \multicolumn{2}{c|}{} & \multicolumn{1}{c|}{}& \multicolumn{1}{c}{\textbf{ (typos-aware)}}\\
			
			\toprule
			Typo type&MRR@10&Recall&MRR@10&Recall&MRR@10&Recall&MRR@10&MRR@10\\
			\toprule
			original             	 &.187&.857&.296&.940&\textbf{.300}&.940&\textbf{.379}&.374\\
			w. typos (avg)	&.120($-35.8\%$)&.696($-18.6\%$)&.141($-52.3\%$)&.712($-24.3\%$)&\textbf{.219${^\dagger}$}($-27.0\%$)&\textbf{.857${^\dagger}$}($-8.8\%$)&.250($-34.0\%$)&\textbf{.289${^\dagger}$}($-22.7\%$)\\
			\midrule
			RandInsert          &.125($-33.1\%$)&.693($-18.9\%$)&.140($-52.7\%$)&.711($-24.4\%$)&\textbf{.225${^\dagger}$}($-25.0\%$)&\textbf{.862${^\dagger}$}($-8.3\%$)&.257($-32.2\%$)&\textbf{.297${^\dagger}$}($-20.6\%$)\\
			RandDelete        &.118($-36.9\%$)&.693($-18.9\%$)&.154($-47.9\%$)&.730($-22.3\%$)&\textbf{.217${^\dagger}$}($-27.6\%$)&\textbf{.853${^\dagger}$}($-9.3\%$)&.257($-32.2\%$)&\textbf{.288${^\dagger}$}($-23.0\%$)\\
			RandSub             &.120($-35.8\%$)&.702($-17.9\%$)&.137($-53.7\%$)&.714($-24.0\%$)&\textbf{.220${^\dagger}$}($-26.7\%$)&\textbf{.858${^\dagger}$}($-8.7\%$)&.250($-34.0\%$)&\textbf{.291${^\dagger}$}($-22.2\%$)\\
			SwapNeighbor  &.122($-34.7\%$)&.702($-17.9\%$)&.137($-53.7\%$)&.705($-25.0\%$)&\textbf{.217${^\dagger}$}($-27.6\%$)&\textbf{.859${^\dagger}$}($-8.6\%$)&.240($-36.7\%$)&\textbf{.284${^\dagger}$}($-24.1\%$)\\
			SwapAdjacent  &.117($-37.4\%$)&.691($-19.1\%$)&.137($-53.7\%$)&.702($-25.3\%$)&\textbf{.214${^\dagger}$}($-28.7\%$)&\textbf{.854${^\dagger}$}($-9.1\%$)&.246($-35.1\%$)&\textbf{.286${^\dagger}$}($-23.5\%$)\\
			\bottomrule
		\end{tabular}
	}
	\label{table:results}
\end{table*}

\section{Results} \label{result}

Empirical results are reported in Table~\ref{table:results}. We note that BM25 if outperformed by both DR and BERT re-ranker across all settings, confirming the superiority of BERT-based methods. For RQ1, we compare the results obtained on the original queries with those on queries with typos, when models are trained using the standard procedure. We observe statistically significant losses in effectiveness for both DR (on average MRR@10 drops $52.3\%$ and Recall  $24.3\%$) and BERT re-ranker (MRR@10 drops $34\%$). BERT re-ranker is performed on top of DR results, thus losses in Recall for DR are propagated to the BERT re-ranker. However, the effectiveness of DR on typos queries drops to about that of BM25, while BERT re-ranker stays superior.



In terms of the impact of different types of typos (RQ2), the results show that different typos have similar impact: they all hurt effectiveness heavily. DR appears however more tolerant to RandDelete typos (with a $\approx5\%$ smaller loss in MRR@10 than for other types of typos), while BERT re-ranker losses are generally uniform across typos types.

To answer RQ3, we compare the results of models produced with typos-aware training vs. with the standard training. 
Despite the typos-aware training, both DR and BERT re-ranker display significant losses in effectiveness when dealing with typos queries, compared to the original queries (Table~\ref{table:results}). 
However, compared to the models with standard training, both methods are much more tolerant to all types of typos when typos-aware training is employed. In fact, losses in MRR@10 halve for DR (from $52.3\%$ to $24.3\%$), and reduce by one third for BERT re-ranker (from $34\%$ to $22.7\%$); all differences are statistically significant. Typos-aware training seems to impact queries with typos produced by RandomInsert more than those with other types of typos. We also note that effectiveness obtained by models with typos-aware training is not different from that with standard training if only queries without typos (original) are considered (minor differences are not statistically significant).

\begin{figure}
	\centering
	\includegraphics[width=1\linewidth]{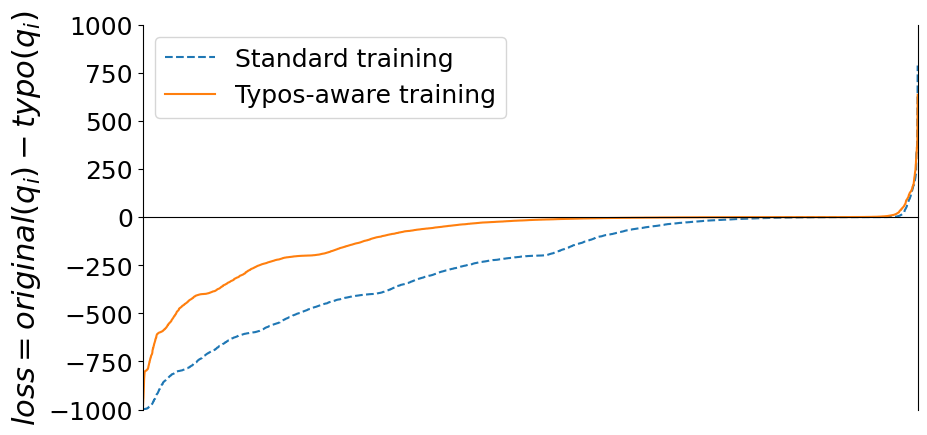}
	\caption{Loss in terms of the rank position of the first relevant passage retrieved by DR when ranking for typos queries, compared to original queries. Each point on the x-axis refers to a query; x-axis ordered by decreasing loss when standard training is used.\vspace{-18pt}}
	\label{fig:gain_loss}
\end{figure}

 
 Figure~\ref{fig:gain_loss} presents the rank loss obtained by DR when answering typos queries in place of the original queries (plot averaged across all types of typos; individual typos types show similar trends). A negative loss of $n$ means when using typos queries, the first relevant document is retrieved $n$ rank positions after that obtained when using the original queries. The figure shows that typos-aware training consistently provides smaller losses than the standard training. We also note there are few cases ($\approx300$ queries) in which typos queries provide gains compared to the original query. Queries with large losses often have typos for keywords that are essential to determine the intent of the query. Typos queries that exhibit gains generally display typos on non-essential keywords, e.g., stopwords.


%% file: sections/case_study.tex
\section{A Case Study} \label{case}
The results presented in the previous sections are conducted with synthetically generated typo queries. Accurate analysis of the MS MARCO dataset revels the presence of a very limit number of queries containing typos -- these typos are legitimate errors made by the user issuing the query. For instance, the MS MARCO dev set contains the mistyped query -- \textit{``sydeny climate''}~\footnote{The correct spelling is ``sydney climate''.} (qid: 506025). Without typos-aware training, the considered DR cannot retrieve the relevant passage in the top 1,000 results. However, with typos-aware training, the DR is able to rank the relevant passage at rank 127 for this typo query. This suggests that DRs trained with our typos-aware training with synthetic typo generation may be able to generalize to real-world typo queries, aside from those synthetic (though realistic) typo queries we considered in our extensive empirical evaluation. In future work, we want to further test our proposed typos-aware training with more real-world typo queries by acquiring a real query log with typos and perform relevance annotations on the MS MARCO passage collection.

%% file: sections/conclusion.tex
\section{Conclusion} \label{conclusion}
In this paper we studied the impact of typos in queries on popular BERT-based passage retrieval methods. We reported significant drops in effectiveness across different types of typos for both DR and BERT re-ranker: these methods are not tolerant to typos in queries when solely relying on the BERT encoder. We then proposed a typos-aware training strategy for DR and BERT re-ranker, which controls the exposure of the models to queries with typos during training. With our typos-aware training, both DR and BERT re-ranker showed to be much more tolerant to typos in queries. We believe our typos-aware training can be used (more extensively than in this paper) as a standard data augmentation step in the DR and BERT re-ranker's training loop since the computations for typos generation are very light and can provide extra gains on typos queries, without hurting effectiveness on queries without typos. Code, typos queries and results files at \url{https://github.com/ielab/typos-aware-BERT}.